\documentclass[prb,twocolumn,aps,showpacs]{revtex4}

\usepackage{amssymb}
\usepackage{amsmath}
\usepackage{graphicx}
\usepackage{dcolumn}
\usepackage{bm}

\newcommand{\vk}{\mathbf{k}}

\begin{document}

\title{Ultrafast demagnetization in the sp-d model: a theoretical study}
\author{{\L}ukasz Cywi\'{n}ski}\email{cywinski@physics.ucsd.edu}
\author{L.~J.~Sham}
\affiliation{Department of Physics, University of California San
Diego, La Jolla, California, 92093-0319}
\date{\today}

\begin{abstract}
We propose and analyze a theoretical model of ultrafast light-induced magnetization dynamics in systems of localized spins that are coupled to carriers' spins by sp-d exchange interaction. A prominent example of a class of materials falling into this category are ferromagnetic (III,Mn)V semiconductors, in which ultrafast demagnetization has been recently observed.
In the proposed model light excitation heats up the population of carriers, taking it out of equilibrium with the localized spins. This triggers the process of energy and angular momentum exchange between the two spin systems, which lasts for the duration of the energy relaxation of the carriers. We derive the Master equation for the density matrix of a  localized spin interacting with the hot carriers and couple it with a phenomenological treatment of the carrier dynamics. 
We develop a general theory within the sp-d model and we apply it  to the ferromagnetic semiconductors, taking into account the valence band structure of these materials.
We show that the fast spin relaxation of the carriers can sustain the flow of polarization between the localized and itinerant spins leading to significant demagnetization of the localized spin system, observed in (III,Mn)V materials. 
\end{abstract}
\maketitle

\section{INTRODUCTION}
During the last decade, ultrafast light-induced magnetization dynamics has been a subject of many experimental studies, which have yielded results of interest from both the insight into the basic physical processes occurring in ferromagnets at short time-scales, and from the possible practical implications for ultrafast writing of magnetic memories. 
The seminal discovery \cite{Beaurepaire_PRL96} of sub-picosecond demagnetization in Ni  has challenged the previously held belief that the fastest time-scale of magnetization dynamics is the spin-lattice relaxation time, which is  of the order of at least tens of picoseconds.\cite{Vaterlaus_PRL91,Hubner_PRB96} Since then, such an ultrafast magnetization quenching process in transition metals has been confirmed by various time-resolved magneto-optical techinques \cite{Beaurepaire_PRB98,Koopmans_PRL00,Bigot_PRL04,Vomir_PRL05}
as well al other experimental methods, such as magnetic second harmonic generation,\cite{Hohlfeld_PRL97} spin-resolved photoemission,\cite{Scholl_PRL97} and THz light emission  \cite{Hilton_JPSJ06}. 
Recently, an analogous phenomenon of light-induced demagnetization, including a complete destruction of ferromagnetic order in less than a picosecond, has been observed in (III,Mn)V ferromagnetic semiconductors.\cite{Wang_PRL05,Wang_JPC06}

Apart from magnetization quenching, which is caused by absorption of light, there has recently been significant progress in nonthermal (coherent) light manipulation of magnetic order. \cite{Kimel_JPC07}
Coherent manipulation on the sub-picosecond time scale has been shown in antiferromagnetic \cite{Kimel_Nature04,Kimel_Nature05} and ferrimagnetic materials.\cite{Hansteen_PRL05}
Theories of coherent manipulation of magnetization by strong off-resonant light have been put forth for (III,Mn)V ferromagnetic semiconductors \cite{Chovan_PRL06} and for closely related undoped paramagnetic (II,Mn)VI materials.\cite{Rossier_PRL04}
In this paper we are interested in ultrafast demagnetization, which is an incoherent process involving strong excitation.  The laser pulse heats up the carrier population, and we analyze the magnetization dynamics induced by such a hot gas of electrons or holes. 

Ultrafast demagnetization experiments in transition metals were initially interpreted using a phenomenological three-temperature model.\cite{Beaurepaire_PRL96,Zhang_TAP02} In this approach the system is divided into three reservoirs: carriers, lattice, and spins. Excitation by light injects energy into the carrier system. Each reservoir is described by its temperature, and phenomenological equations are written down for heat flow between each pair of reservoirs. Physical underpinnings and time-scales of two of the couplings, carrier-lattice (electron-phonon) and spin-lattice interactions are known (for the latter see Ref.~\onlinecite{Hubner_PRB96}). A new ingredient, a direct  carrier-spin coupling has to be postulated in order to  explain the ultrafast demagnetization. In addition to the lack of detailed microscopic understanding of such carrier-spin coupling, there are two major deficiencies of the three-temperature model. The first is that treating carriers and spins as separate entities might not be a good starting point for the transition-metal itinerant ferromagnets. If such an approach is possible, the nature of an effective separation into subsystems should be elucidated. The second shortcoming is the fact that only the energy transfer between reservoirs is considered. As the key phenomenon to be explained is the change of magnetization, a correct physical description should involve the mechanism by which the angular momentum (spin and orbital) is exchanged between the subsystems.\cite{Koopmans_TAP03,Koopmans_JMMM05}
Recently,\cite{Koopmans_PRL05} Koopmans et.~al.~proposed a  simplified model of a transition-metal ferromagnet in which spinless electrons induce spin-flips in a magnetic subsystem separate from the electronic system. 
Another proposed theory of demagnetization,\cite{Zhang_PRL00} in which the cooperative effect of a coherent laser field and the spin-orbit coupling is calculated, seems to be more suitable for the case of coherent manipulation of antiferromagnets.\cite{Gomez-Abal_PRL04,Kimel_JPC07}

A model of ferromagnetism ideally suited for investigation of magnetization quenching induced by excitation of carriers is the sp-d model. 
In this approach, most of the macroscopic magnetization comes from the localized d-shell spins (or f shells, in case of the rare earths), which are coupled by an exchange interaction to itinerant s or p carriers. 
This model was introduced independently by Zener \cite{Zener_PR51} and by Vonsovskii \cite{Vonsovskii} in order to describe the transition metals, but was abandoned when the above-mentioned lack of sharp separation into s carriers and d spins was understood. Later,  the s-d(f) model was succesfully applied to magnetic semiconductors such as the chalcogenides of europium.\cite{Nagaev}
Recently, it has been again used in itinerant ferromagnets in order to analyze the situations in which an interplay between the transport and magnetic properties is critical. Examples include spin-transfer torque  \cite{Slonczewski_JMMM96,Berger_PRB96}  and an enhancement of Gilbert damping in a magnet due to pumping of spin currents into adjacent non-magnetic material.\cite{Tserkovnyak_RMP05,Simanek_PRB03}
Finally, and most importantly for us, the p-d model is used to describe the basic physics of ferromagnetism in (III,Mn)V semiconductors,\cite{Dietl_Science00,Jungwirth_RMP06} such as Ga$_{1-x}$Mn$_{x}$As or In$_{1-x}$Mn$_{x}$As .

Ferromagnetic (III,Mn)V semiconductors are created by doping a small ($\sim$5\%) molar fraction of Mn into a III-V semiconductor such as GaAs or InAs. The Mn ions, which substitute the cations, are acceptors and their d shells can be treated as well-localized $5/2$ spins.\cite{Dietl_SST02} Hybridization between Mn d orbitals and anion p states leads to a kinetic p-d exchange interaction \cite{Kacman_SST01}, which couples localized d spins to spins of p holes in the valence band. These holes are a source of indirect Mn-Mn spin coupling. At large hole densities (typically $p$$\approx$$10^{20}$cm$^{-3}$) a mean-field theory of Zener \cite{Zener_PR51,Dietl_Science00,Dietl_PRB01} has been successful in describing many features of ferromagnetism in these materials. It correctly predicts the critical temperature $T_{c}$, increasing trend in $T_{c}$ with the density of carriers, and magnetic anisotropies.\cite{Jungwirth_RMP06}

The strong correlation between the presence of the delocalized (or at least weakly localized) holes and ferromagnetic ordering of Mn spins, has been firmly established experimentally. 
The optical induction of ferromagnetic transition (through cw photoinjection of holes) has been shown,\cite{Koshihara_PRL97} and the critical temperature and the coercive field have been altered by changing the density of holes in InMnAs by applying a gate voltage.\cite{Ohno_Nature00,Chiba_Science03}
Most relevant for this article is the observation of the sub-picosecond light induced demagnetization in InMnAs \cite{Wang_PRL05} and GaMnAs.\cite{Wang_JPC06} A complete quenching of ferromagnetic order was achieved in InMnAs for pump fluences above 10 mJ/cm$^{2}$. This should be contrasted with behavior in Ni, where the demagnetization saturates at higher fluences while not reaching the complete demagnetization.\cite{Cheskis_PRB05} Since the carrier concentration is much lower in ferromagnetic semiconductors as compared to metals, their magnetization is  more amenable to manipulation by external stimuli.  

These experiments led us to a preliminary investigation of the spin dynamics induced by strong incoherent excitation within the framework of the sp-d model,\cite{Wang_PRL05} and to a proposal of the ``inverse Overhauser effect'' as the physical basis of the observed phenomenon.  
In this scenario, the rate of spin-flip scattering between the localized spins and the mobile carriers is  enhanced by the non-equilibrium distribution of excited carriers. This triggers the transfer of angular momentum from the localized spin system to the carriers, leading to demagnetization of the localized spins. For this process to be sustained, the angular momentum transferred into the carriers' system has to be efficiently dissipated into the lattice by spin-orbit interaction assisted scattering, and we have stressed that the expected short spin relaxation time of holes in ferromagnetic semiconductor is essential for the explanation of measured changes in magnetization.

In this article, we present a detailed theory of ultrafast demagnetization within the sp-d model.
We concentrate on ferromagnetic semiconductors, specifically, we discuss photoexcitation processes specific to these materials, and we perform calculations using an effective Hamiltonian \cite{Jungwirth_RMP06} model of the spin-split valence band. 
We assume that the indirect (carrier-mediated) exchange interaction between localized spins dominates over short-range antiferromagnetic d-d superexchange (it has been argued\cite{Dietl_PRB01} that in the presence of holes the superexchange is suppressed). We also neglect the Mn interstitials\cite{Yu_PRB02} which can form antiferromagnetically coupled pairs with nearby substitutional Mn spins,\cite{Blinowski_PRB03} as we are only interested in Mn spins which participate in ferromagnetic order. Furthermore we work in the regime of strong carrier excitation, in which the carrier mediated Mn-Mn correlations beyond the mean-field level are averaged out. The localized spins flip independently in the common mean field due to the average carrier spin.
We derive the rate equations for their dynamics due to spin-flip scattering with the carriers spins. The transition rates depend on the instantaneous state of the carrier system at a given time. 
This derivation generalizes to the non-stationary case the theory from Ref.~\onlinecite{Konig_PRB00}, where the heating of Mn spins by electrons excited by cw light was considered in a (II,Mn)VI based quantum well. Apart from the localized spin dynamics, we consider the energy and spin relaxation of carriers by means of phenomenological equations. The energy relaxation time of the carriers sets an upper bound for the time-scale of the ultrafast demagnetization process. If the spin relaxation rate of carriers is smaller than the rate at which the angular momentum is injected into the carrier system by sp-d scattering, the demagnetization becomes suppressed. Below we show that this ``spin bottleneck'' effect does not affect qualitatively the results in p-type (III,Mn)V semiconductors.

The general results of this article are directly applicable to any system described by analogous s(p)-d(f) model.
However, when the non-carrier-mediated d-d exchange coupling is strong (e.g. in the europium chalcogenides \cite{Nagaev}), the starting point of the calculation should not be the interaction of a single spin with the carriers (which is our approach here), but the enhancement of the carrier-magnon scattering. A similar treatment should be applied to a recent measurement\cite{Wang_preprint07} of the ultrafast demagnetization in GaMnAs excited with the pump fluence 3 orders of magnitude smaller than in Ref.~\onlinecite{Wang_PRL05}. In this case, our assumption of complete obliteration of the carrier-mediated exchange coupling by excitation of carriers might not hold, and spin-wave modes of coupled Mn spins should be the starting point of the calculation.

The article is organized as follows:  
In section \ref{sec:sp-d} we introduce the sp-d model. Section \ref{sec:excitation} contains a discussion of the photoexcitation process, which is followed in Section \ref{sec:bath} by the description of the model of the carrier bath heated up by absorption of light.
Section \ref{sec:derivation} contains the formalism and the derivations  of equations governing the dynamics of localized spins interacting with hot itinerant spins.
In Section \ref{sec:1band} we calculate the process of ultrafast demagnetization using a simplified band structure (a single spin-split band), introduce the equations governing the carrier dynamics, and discuss how the energy and spin relaxation of carriers influences the demagnetization process. Complications introduced by valence band-structure of a ferromagnetic semiconductor are discussed in Section  \ref{sec:valence}, where we use a 6 band Luttinger Hamiltonian to calculate the hole-Mn spin-flip transition rate.
Finally, we discuss the connection to the experiments in Section \ref{sec:comparison}.

\section{GENERAL THEORY}
\subsection{The sp-d model} \label{sec:sp-d}
We use a Hamiltonian consisting of the single-particle carrier part $\hat{H}_{C}$, localized spin part $\hat{H}_{S}$ and the sp-d coupling $\hat{H}_{CS}$: 
\begin{equation}
\hat{H} = \hat{H}_{C} + \hat{H}_{S} + \hat{H}_{CS}  \,\, .
\end{equation}
We do not explicitly consider the carrier-carrier interaction, carrier-phonon interaction, and the lattice dynamics. However, these interactions are included phenomenologically in our treatment of the carrier bath (Section \ref{sec:bath}).
The carrier part of the Hamiltonian is given by
\begin{equation}
\hat{H}_{C} = \sum_{n\mathbf{k}} \epsilon_{n\mathbf{k}} {a^{\dagger}}_{\!n\mathbf{k}} a_{n\mathbf{k}\sigma} \,\, ,  \label{eq:HC}
\end{equation}
where $n$ is the band index, $\mathbf{k}$ is the wave-vector, and $\epsilon_{n\vk}$ is the energy. In the case of (III,Mn)V semiconductors, the valence band structure near the $\Gamma$ point can be obtained from a $\vk\cdot\mathbf{p}$ model, such as 6$\times$6 Luttinger Hamiltonian commonly used in these systems.\cite{Dietl_PRB01} Although the ferromagnetic semiconductors are known to be heavily disordered, we have neglected the disorder potential in Eq.~(\ref{eq:HC}). 
If the disorder can be treated perturbatively, then its influence on the treatment below is not expected to be strong. As we are working in the regime of very strong excitation of carriers, more subtle features of bands will be averaged out. 
On the other hand, if the effects of disorder are non-perturbative (e.g. the holes are in an impurity band rather than in the host valence band, as it was recently suggested in Ref.~\onlinecite{Burch_PRL06}), a new energy dispersion (e.g. a single weakly dispersive band) should be included introduced in Eq.~(\ref{eq:HC}). 

The localized spin Hamiltonian $\hat{H}_{S}$ includes Zeeman coupling to the external magnetic field.
In the ferromagnetic phase, the splitting of the localized spin S due to exchange interaction with carriers exceeds the typical Zeeman splitting, and the only role of Zeeman term is to choose the direction of the magnetization. We disregard possible S-S exchange coupling by mechanisms other\cite{Kacman_SST01} than indirect carrier-mediated exchange.
 
Finally, the exchange coupling is given by
\begin{equation}
\hat{H}_{CS} = -\frac{\gamma}{V} \sum_{l} \!\! \sum_{n\mathbf{k}, n'\mathbf{k}'} \!\!\mathbf{\hat{S}}_{l}\cdot \langle n\vk| \mathbf{\hat{s}} | n'\vk'\rangle   \,   e^{i(\mathbf{k}'-\mathbf{k})\mathbf{R}_{l}} \,\, {a^{\dagger}}_{\!n\mathbf{k}} a_{n'\mathbf{k}'}   \label{eq:HCS} \,\, ,
\end{equation}
where $\mathbf{\hat{S}}_{l}$ is the spin operator of localized spin at $\mathbf{R}_{l}$, $\mathbf{\hat{s}}$ is the carrier spin operator and $\gamma$ is the exchange constant.  In the literature on diluted magnetic semiconductors, the $\gamma$ parameter  is called $\alpha$ and $\beta$ for the conduction and valence band electrons, respectively.\cite{Furdyna_JAP88} The typical values in (III,Mn)V materials are $\alpha$$\approx$$10$ and  $\beta$$\approx$$-50$ meV$\cdot$nm$^{3}$.  The exchange energy $J$ per unit cell is $J$$=$$N_{0}\gamma$, where $N_{0}$ is the density of the cations. Accordingly, $N_{0}\alpha$$\approx$$0.2$ eV and $N_{0}\beta$$\approx$$-1$ eV.

For a single band, the sum over $n$ states is simplified to
\begin{equation}
\hat{H}^{1}_{CS} =  -\frac{\gamma}{V} \sum_{l} \sum_{\mathbf{k}, \mathbf{k}'}  \mathbf{\hat{S}}_{l}\cdot \mathbf{\hat{s}}_{\vk\vk'}    \,   e^{i(\mathbf{k}'-\mathbf{k})\mathbf{R}_{l}}  \label{eq:HCS1}\,\, ,
\end{equation}
where the scalar product of spin operators can be expressed using spin ladder operators as: 
\begin{equation}
\mathbf{\hat{S}}_{l}\cdot \mathbf{\hat{s}}_{\vk\vk'} = \frac{1}{2} ( \hat{S}_{l}^{+}\hat{s}^{-}_{\vk\vk'} + \hat{S}_{l}^{-}\hat{s}^{+}_{\vk\vk'} ) + \hat{S}_{l}^{z}\hat{s}^{z}_{\vk\vk'}  \,\, ,
\end{equation}
and the explicit form of carrier spin operators is
\begin{equation}
\hat{s}^{z}_{\mathbf{k}\mathbf{k}'} = \frac{1}{2} ( {a^{\dagger}}_{\!\mathbf{k}+} a_{\mathbf{k}'+} - {a^{\dagger}}_{\mathbf{k}-} a_{\mathbf{k}'-} )
\end{equation}
\begin{equation}
\hat{s}^{+}_{\mathbf{k}\mathbf{k}'} =  {a^{\dagger}}_{\!\mathbf{k}+} a_{\mathbf{k}'-} \,\,\, , \,\,\, \hat{s}^{-}_{\mathbf{k}\mathbf{k}'} =  {a^{\dagger}}_{\!\mathbf{k}-} a_{\mathbf{k}'+}  \,\, .\end{equation}

The magnetization (average localized spin) is in the $z$ direction, and we treat the sp-d Hamiltonian in a mean-field approximation. The mean-field felt by the carriers is obtained by performing the simplest virtual crystal disorder averaging of localized spins positions: 
\begin{equation}
\hat{H}^{mf}_{C}  =  -n_{i}\gamma \langle S_{z} \rangle \sum_{n,n',\mathbf{k}}  \langle n\vk| \hat{s}^{z} | n'\vk\rangle {a^{\dagger}}_{\!n\mathbf{k}} a_{n'\mathbf{k}} \,\, ,  \label{eq:Cmf}
\end{equation}
where $n_{i}$ is the density of localized spins and $\langle S_{z} \rangle$ is the average localized spin. The energy of the carrier spin splitting is defined as
\begin{equation}
\Delta =  -n_{i}\gamma \langle S_{z} \rangle \,\, . \label{eq:Delta}
\end{equation}
The typical value of $\Delta$ in GaMnAs and InMnAs with the highest critical temperatures is about $0.1$ eV. 
By averaging Eq.~(\ref{eq:HCS}) with respect to the carriers we arrive at
\begin{equation}
\hat{H}^{mf}_{S} = \delta \sum_{l} \hat{S}^{z}_{l}  \,\, ,  \label{eq:Smf}
\end{equation}
where the energy splitting of different $m$ levels of a localized spin is 
\begin{equation}
\delta = - n_{c}\gamma \langle s_{z} \rangle  \,\, , \label{eq:delta}
\end{equation}
with $n_{c}$ the carrier density and $\langle s_{z} \rangle$ the average carrier spin. The typical magnitude of $\delta$ in (III,Mn)V materials is of the order of 1 meV. In the mean-field theory of ferromagnetism \cite{Dietl_Science00,Dietl_PRB01,Rossier_PRB01} only the two averaged fields $\langle S_{z} \rangle$ and $\langle s_{z} \rangle$ are taken into account, and the free energy of their interaction is minimized.

\subsection{Light excitation of the carrier system in (III,Mn)V} \label{sec:excitation}
\begin{figure}
\includegraphics[width=8cm]{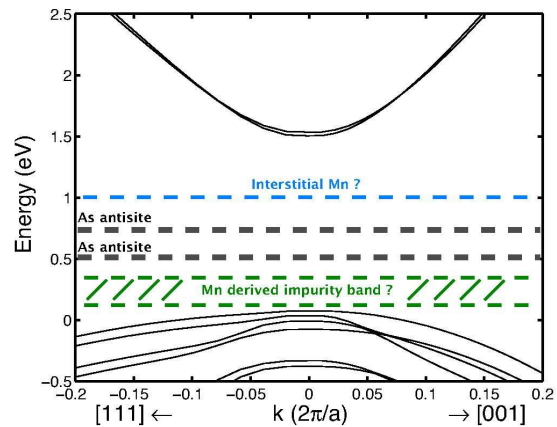}
\caption{(Color online) The band structure of GaMnAs (solid lines) calculated using 8 band {$\mathbf{k}\cdot\mathbf{p}$} model with mean-field sp-d exchange interaction ($|\Delta|$$=$$0.15$ eV). The dotted lines denote the possible positions of midgap energy levels of different origins. In the model without an impurity band, the Fermi energy is typically around $-0.1$ eV for hole density $p$$\sim$$10^{20}$ cm$^{-3}$. } \label{fig:GaMnAs_bandstructure}
\end{figure}

To construct a model of the excited carrier bath, it is helpful to first analyze qualitatively the process of carrier photoexcitation in ferromagnetic semiconductors. The cw (magneto)optical spectra of GaMnAs are qualitatively different from the ones in pure GaAs, as they show very strong effects of disorder.\cite{Szczytko_PRB01,Komori_PRB03,Singley_PRB03,Lang_PRB05} In GaMnAs, there is no gap in the absorption, which is at least of the order of $10^{4}$ cm$^{-1}$ for all energies.\cite{Singley_PRB03} The origin of strong absorption inside the host gap is a matter of controversy (see Figure \ref{fig:GaMnAs_bandstructure}). It has been proposed that the additional optical transitions are between the valence band and the dispersionless levels (located about $0.7$ E$_{g}$ above the valence band) originating from As antisites or Mn interstitials.\cite{Hankiewicz_PRB04}  
Transitions terminating inside the Mn derived impurity band $\sim$$0.1$ eV above the valence band edge have also been suggested.\cite{Hwang_PRB02,Singley_PRB03,Burch_PRL06}

In addition, the role of the inter-valence band transitions is increased in the disordered material, in which the $\mathbf{k}$-selection rule is relaxed. As a result, the absorption of light with energy smaller than the bandgap of the host material leads to a strong excitation in the valence band, which occurs due to inter-valence transitions and possible transitions from below the Fermi energy into the localized states within the gap. The initial distribution of photoholes is expected to be very broad, determined only by energy conservation (not by $\mathbf{k}$-selection).
We use the broadness of the distribution of the carriers after the excitation and their total number as free parameters of the theory. Some flexibility in the used values of total hole density immediately after the excitation $p$ is also justified by the fact that the initial densities $p_0$ are usually known with only an  order of magnitude accuracy.
We use an energy width of hole distribution after excitation of the order of $0.1$ eV, a value comparable to the exchange splitting of the valence band. In Section \ref{sec:comparison} we use the experimental data to argue that the above energy scale is sensible.

The optical experiments in InMnAs are limited to samples with very small Mn concentrations \cite{Hirakawa_PE01} or energies above the fundamental bandgap.\cite{Fumagalli_PRB96} The role of various defects has not been investigated in InMnAs as closely as in \nobreak{GaMnAs}, but the observed\cite{Wang_JPC06} ultrafast electron trapping time suggests that InMnAs does have a large concentration of midgap defects characteristic of low-temperature grown III-V semiconductors.
We can also use the results of this article to argue for the presence of strong transitions involving the valence band states and states not in the conduction band, on the ground that the number of states available in the conduction band is far too small to explain the demagnetization results.

This picture of photoexcitation is confirmed by  the ultrafast demagnetization measurements \cite{Wang_JPC06} in GaMnAs excited by $0.6$ eV pump, far below $1.5$ eV bandgap of GaAs . The results are similar to the ones observed \cite{Wang_PRL05} in InMnAs (bandgap of $0.4$ eV) excited by the same pump beam, showing that excitation into the conduction band plays a minor role in the demagnetization process.

\subsection{Model of the carrier bath} \label{sec:bath}
According to the discussion in the previous section, immediately after the photoexcitation the distribution of carriers is very broad. We will approximate it by a thermal distribution described by the electronic (carrier) temperature $T_{\text{e}}$. 
Using the experimental data from Ref.~\onlinecite{Wang_PRL05} we estimate the initial  $T_{\text{e}}$ to be of the order of 1000 K (for details see Section \ref{sec:comparison}).

The interaction between the carriers and the localized spins produces a mean-field term (Eq.~(\ref{eq:Cmf})) and a secondary term corresponding to simultaneous flips of the itinerant and localized spins. The latter causes exchange of angular momentum between the excited carriers and the localized spin system. Due to the spherical symmetry of the sp-d Hamiltonian, the sp-d interaction alone conserves the total spin, and it can only move the spin polarization from one system to another.
The mechanism of this transfer of spin is described in detail in the next Section. Now we concentrate on the features of the carrier bath which are specific to the case at hand: the possibility of dynamic spin polarization and the energy relaxation of carriers.

The spin transferred into the carrier system is not conserved due to the spin-orbit interaction.  In its presence the scattering within the electronic system is accompanied by spin relaxation \cite{Zutic_RMP04} (ultimately into the lattice). However, the spin relaxation occurs on a finite time-scale $\tau_{\text{sr}}$.
If the rate at which the carriers relax their spins is smaller than the rate at which angular momentum is injected into the carrier population, there is a dynamic spin polarization of the carriers. In this case the average carrier spin deviates  from the mean-field value determined by $\Delta$ splitting of the bands.
In the case when the carriers occupy a single spin-split band, there is a simple way of introducing the dynamic polarization into our formalism 
(the case of multiple bands strongly coupled by spin-orbit interaction is considered in Section \ref{sec:valence}).
We assume that populations for both spin directions are described by  Fermi-Dirac distributions with the common temperature $T_{\text{e}}$ but different quasi-chemical potentials $\mu_{s}$ (spin $s$$=$$\uparrow$,$\downarrow$). 
This is equivalent to saying that we coarse-grain the dynamics on a time-scale larger than the time in which the energy is redistributed within the carrier system (by carrier-carrier scattering), and we explicitly consider only the slower processes: spin relaxation of carriers and energy transfer into the lattice.
It is similar to the situation encountered in semiconductor lasers, where the processes of thermalization of electrons and holes separately occur faster than the recombination, and the resulting globally out-of-equilibrium situation can be described by using the quasi-equilibrium form of the carrier distribution,  with different quasi-chemical potentials for electrons and holes. 

As we discuss in detail in the following sections, dynamic polarization of carriers affects the spin-flip scattering rate. This is a back-action effect of the S system on the carrier system: demagnetization of localized spins causes the dynamic polarization of carriers, which in turn influences the rate of transfer of angular momentum between the systems. Additionally, as the average S spin changes in time, so does the $\Delta$ splitting of the carrier bands. 

Apart from the dynamic polarization effects, we also have to take into account that the carriers are not in thermal equlibrium with the lattice, and in the first picoseconds after the photoexcitation they are described by a different temperature than the lattice. We model the carrier-phonon interaction (leading to the cooling of carriers) and carrier-carrier scattering (maintaining the thermal distribution) phenomenologically. In metals, the electron-phonon energy relaxation time is a couple hundreds of femtoseconds.\cite{Groeneveld_PRB95,VanKampen_JPC05}
In semiconductors, the regime of such strong excitation as used in Ref.~\onlinecite{Wang_PRL05} to demagnetize InMnAs has not been investigated in detail. 
Theoretical calculations of energy relaxation of holes deep in the valence band indicate that emission of optical phonons is very efficient for these states.\cite{Woerner_PRB95} Calculated emission of 7 phonons during a picosecond should correspond to significant cooling of the carriers. Na{\"i}ve extrapolation of energy-loss curves \cite{Shah} to hole temperatures of $\sim$$1000$K also gives a sub-picosecond energy relaxation time. In the following, we will use an energy relaxation time $\tau_{\text{E}}$ of the order of picosecond, and assume $T_{\text{e}}(t)=T_{\text{e}}(0)\exp(-t/\tau_{\text{E}})$. More generally, the carrier and lattice temperatures could be solved for using a two-temperature model.\cite{Allen_PRL87} However, in the case of (III,Mn)V semiconductors the rise in lattice temperature due to heat transfer from the carriers described by such $T_{\text{e}}$ is quite small, so that the final common temperature of carriers and the lattice is less than 100 K. As we discuss in detail further on, such a temperature corresponds to very slow demagnetization dynamics, and for the purpose of calculating the ultrafast process, we can then assume that the carrier temperature simply decays  towards zero. 

In the following, all the averages with respect to carrier degrees of freedom will be taken using a density matrix of non-interacting electrons $\hat{\rho}_{C}$, which is not necessarily of the equilibrium form, but remains diagonal in the basis of carrier Hamiltonian's eigenstates $| \alpha \rangle$$=$$| n \vk \rangle$.
These averages are denoted as $ \langle ... \rangle_{C}$$\equiv$$\text{Tr}_{C}\{ \hat{\rho}_{C}...\}$, and they are given by
\begin{eqnarray}
\langle {a^{\dagger}}_{\alpha}a_{\beta} \rangle_{C}  & = & \delta_{\alpha\beta} \,\,  f_{\alpha} \\
\langle {a^{\dagger}}_{\alpha} a_{\beta} {a^{\dagger}}_{\gamma} a_{\delta} \rangle_{C}  & = & \delta_{\alpha\beta}\delta_{\gamma\delta}  \,\, f_{\alpha} f_{\gamma} \\
& & + \delta_{\alpha\delta}\delta_{\beta\gamma} \,\, f_{\alpha}(1-f_{\beta}) \nonumber
\end{eqnarray}
where $f_{\alpha}$ is the average occupation of $\alpha$ state. In a single-band model, the $\alpha...\delta$ indices refer to $| s \vk \rangle$ states with spin $s$$=$$\uparrow$,$\downarrow$. $f_{s\vk}$  is a Fermi-Dirac function at temperature $T_{\text{e}}$ with spin-dependent chemical potential $\mu_{s}$. 

\subsection{Rate equations for the localized spin} \label{sec:derivation}
The framework of the sp-d model allows for clear separation of carrier (C) and localized spin (S) systems. The mean-field parts of their mutual interaction are 
given by Eq. (\ref{eq:Cmf}) and (\ref{eq:Smf}). The spin-splitting $\Delta$ of the carriers' band is proportional to the instantaneous $\langle S_{z} \rangle$, and the splitting of localized spins, $\delta$, changes with the average carrier spin $\langle s_{z} \rangle$.
We assume that any correlation between localized spins beyond the mean-field Zener approach is obliterated by the strong scattering of excited carriers.
Each localized spin feels the dynamics of the other spins only through their average value, which influences the state of the carrier system (its spin splitting $\Delta$). Essentially, we consider an ensemble of paramagnetic S spins interacting with a bath, the properties of which depend on the average S. The ferromagnetism enters only as an initial condition: the S system is polarized at $t$$=$$0$. Below we derive the equations for the dynamics of the average S spin due to the interaction with the carrier bath, the state of which depends on the average S.

The Hamiltonian of the localized S spins and the carriers system is written as
\begin{equation}
\hat{H}_{S-C} = \hat{H}_{0}+\hat{V} = \hat{H}_{C} + \hat{H}^{mf}_{C} + \sum_{l}( \delta \hat{S}^{z}_{l} + \hat{V}_{l} ) \,\, ,  \label{eq:HS-C}
\end{equation}
where $\hat{H}_{C}$ is the carrier band Hamiltonian  (Eq.~(\ref{eq:HC})), $\hat{H}^{mf}_{C}$ is the mean-field spin-splitting from Eq.~(\ref{eq:Cmf}), $\delta$ is the mean-field localized spin splitting defined in Eq.~(\ref{eq:delta}), and  the spin-flip term of $l$-th localized spin $\hat{V_{l}}$ comes from part of sp-d interaction which is off-diagonal in $\hat{S}^{z}$ basis. We write it in the following way:
\begin{equation}
\hat{V}_{l} = \hat{S}^{+}_{l}\hat{F}^{-} + \hat{S}_{l}^{-}\hat{F}^{+}	\label{eq:Hsf}
\end{equation}
where $\hat{F}^{\pm}$ are proportional to the ladder operators of carrier spin. In the general case of multiple bands (as in Eq.~(\ref{eq:HCS})) we have
\begin{equation}
\hat{F}^{\pm} = -\frac{\gamma}{2V} \sum_{n\mathbf{k}, n'\mathbf{k}'} \langle n\vk| \mathbf{\hat{s}^{\pm}} | n'\vk'\rangle  {a^{\dagger}}_{\!n\mathbf{k}} a_{n'\mathbf{k}'} \,\, ,  \label{eq:Fn}
\end{equation}
whereas for a single band we have
\begin{equation}
\hat{F}^{\pm} =  -\frac{\gamma}{2V} \sum_{\vk, \vk'} {a^{\dagger}}_{\!\mathbf{k}\pm} a_{\mathbf{k}'\mp} \,\, .  \label{eq:F1}
\end{equation}

Now we follow a standard way of deriving the Master equation for the density matrix of the localized spin system interacting with a carrier bath (see e.g.~Ref.~\onlinecite{Blum}). The total density matrix of the system is assumed to factorize into the carrier and the localized spin density operators: 
\begin{equation}
\hat{\rho}(t) \approx \hat{\rho}_{C}(t) \hat{\rho}_{S}(t) \,\, , \label{eq:irreversibility}
\end{equation}
and the Liouville equation for time-dependence of $\hat{\rho}(t)$ is turned into an equation for  $ \hat{\rho}_{S}(t)$ by tracing out the carrier degrees of freedom. 
However, unlike in standard treatment,\cite{Blum} the state of the carrier bath changes in time, as discussed in the previous section. 

The usual derivation of the Master equation implies coarse-graining of the system dynamics on time-scale $\Delta t$ longer than the correlation time of the bath, $\tau_{c}$. This is the condition on which the Markov approximation rests.
For the gas of carriers described by the effective temperature $k_{B}T_{\text{e}}$$\approx$$0.1$ eV, we can expect this time to be of the order of a few femtoseconds.  Let us define a time-scale $\tau_{\rho}$ in which changes in  the carriers' density matrix $\hat{\rho}_{C}(t)$ occur. The contributions to the evolution of $\hat{\rho}_{C}(t)$ are as follows. The carrier temperature changes appreciably during energy relaxation time $\tau_{\text{E}}$$\sim$$1$ ps. The spin splitting of the band $\Delta(t)$ is proportional to the average localized spin $\langle S(t) \rangle$, which we expect to decrease during a characteristic demagnetization time $\tau_{\text{M}}$. The build-up of the dynamic spin polarization of the carriers is determined by two processes: transfer of spin from the S system occurring during the aforementioned time $\tau_{\text{M}}$ and the spin relaxation of carriers characterized by time $\tau_{\text{sr}}$ (for very short $\tau_{\text{sr}}$, there is no dynamical polarization). Now, we will assume that all these result in the time $\tau_{\rho}$ much larger than the bath correlation time $\tau_{c}$, so that we can choose our coarse-graining step $\Delta$ fulfilling the condition:
\begin{equation}
\tau_{c} \ll \Delta t \ll \tau_{\rho}  \,\, .
\end{equation}
In such a case, at each coarse-grained time-step $t_{n}$ we can derive a Master equation with carriers described by $\hat{\rho}_{C}(t_{n})$ treated as approximately constant during $\Delta t$. In this way we can use the Markov approximation locally in time, having separated the ``macroscopic'' back-action of the localized spin system on the bath, which occurs on a longer time-scale. The Master equation for the localized spin density matrix $\hat{\rho}_{S}$ is then derived exactly as in the usual case (see Ref.~\onlinecite{Blum}), only with the transition rates depending on time $t_{n}$. We write the equations in the continuum limit, keeping in mind that they cannot be used for times shorter than $\Delta t$, thus obtaining  the following rate equations for the diagonal elements of the localized spin density matrix $\rho^{S}_{m}$$\equiv$$\rho^{S}_{m,m}$:
\begin{eqnarray}
\frac{d}{dt} \rho^{S}_{m} & = & -( W_{m-1,m} + W_{m+1,m}) \rho^{S}_{m} \nonumber \\
& & + W_{m,m+1}\rho^{S}_{m+1} + W_{m,m-1}\rho^{S}_{m-1} \,\, , \label{eq:rate}
\end{eqnarray}
where $W_{n,m}$ is the transition rate from $m$ to $n$ energy  level of the localized spin induced by sp-d interaction with the carriers. The time dependence of the transition rates is suppressed for clarity.
Then the average localized spin evolves according to
\begin{equation}
\frac{d}{dt} \langle S_{z}(t) \rangle = \sum_{m} m\frac{d}{dt}\rho^{S}_{m} \,\, . \label{eq:general_dSdt}
\end{equation} 

The general formula for $W_{m,m\pm1}$ is
\begin{equation}
W_{m,m\pm 1}(t) = \frac{1}{\hbar^2} S^{\mp}_{m,m\pm1}  \int_{-\infty}^{\infty} \!\!\! dt'  \, e^{\pm i\delta t'/\hbar} C_{\mp\pm}(t;t') \,\, , \label{eq:WC}
\end{equation}
in which the matrix element (squared) for the flip of the localized spin is given by
\begin{equation}
S^{\mp}_{m,m\pm1} =  S(S+1)-m(m\pm 1)  \,\, ,
\end{equation}
where $S$ is the magnitude of the localized spin. 
The correlation function of the carriers $C_{ij}(t;t')$, with $i$,$j$$=$$\pm$, is given by (reverting to the coarse-grained notation with $t_{n}$ replacing t)
\begin{equation}
C_{ij}(t_{n};t') = \text{Tr}_{C} \{ \hat{\rho}_{C}(t_{n}) \tilde{F}_{n}^{i}(t_{n}+t')  \tilde{F}_{n}^{j}(t_{n}) \} \,\, ,
\end{equation}
where $\text{Tr}_{C} \{ ...\}$ is the trace with respect to the carrier states and  $\tilde{F}_{n}^{i}(t)$ are operators defined in Eqs.~(\ref{eq:Fn}) and (\ref{eq:F1}) written in the ``local'' interaction picture at coarse-grained time $t_{n}$:
\begin{equation}
\tilde{F}_{n}^{i}(t'') =  \exp \left\{ \frac{i}{\hbar} \hat{H}_{0}(t_{n}) t'' \right\}  \hat{F}^{i}(t'') \exp \left\{-\frac{i}{\hbar} \hat{H}_{0}(t_{n}) t'' \right\}  \,\, ,
\end{equation}
where $\hat{F}^{i}(t)$ is the operator in the Schr{\"o}dringer picture, and $\hat{H}_{0}(t_{n})$, defined in Eq.~(\ref{eq:HS-C}), depends on time through the mean-field spin splittings $\delta$ and $\Delta$.
These correlation functions $C_{ij}(t_{n};t')$ decay for $t'$ larger than the correlation time $\tau_{c}$, which has to be much shorter than the time on which $\hat{\rho}_{C}(t)$ changes. For this reason the integration domain in Eq.~(\ref{eq:WC}) is effectively $t'\in(-\tau_{c},\tau_{c})$. In this range of $t'$, the above definitions of $C_{ij}$ and $\hat{F}_{n}^{i}(t)$ make sense. 

In the general case of the multiple bands the transition rates are given by:
\begin{widetext}
\begin{equation}
W_{m,m\pm 1}  =  \frac{\gamma^{2}}{4} \frac{2 \pi}{\hbar} S^{\mp}_{m,m\pm1}
\sum_{nn'}  \int  \frac{\text{d}^{3}k}{(2\pi)^{3}} \int \frac{\text{d}^{3}k'}{(2\pi)^{3}}  | \langle n'\vk'| \hat{s}^{\pm} | n\vk \rangle |^{2}  f_{n\vk}(1-f_{n'\vk'}) \, \delta( \tilde{\epsilon}_{n\vk} - \tilde{\epsilon}_{n'\vk'} \pm \delta ) \,\, ,\label{eq:Wmn}
\end{equation}
\end{widetext}
where $n$ and $n'$ are labeling the subbands, $f_{n\vk}$ is the occupation of $|n\vk\rangle$ state, and $\tilde{\epsilon}_{n\vk}$ are the band energies with the mean-field exchange interaction with localized spins taken into account. The distribution functions, energies  $\tilde{\epsilon}_{n\vk}$, and $\delta$ depend implicitly on time, as we discussed before.

For a single spin-split band we replace $n$ and $n'$ by two spin indices and recover the formula given in Ref.~\onlinecite{Wang_PRL05}. An analogous expression has been used in Ref.~\onlinecite{Konig_PRB00}, where heating of the Mn spins by photoelectrons was considered in a paramagnetic (II,Mn)VI quantum well. Actually, in the case of carriers being a true reservoir of energy and polarization, Eqs.~(\ref{eq:rate}) and (\ref{eq:Wmn}) were derived originally by Korringa\cite{Korringa_Physica50} in order to describe the relaxation of nuclear spins coupled to carriers' spins by hyperfine interaction. 

The rate equation (\ref{eq:rate}) has to be augmented by equations governing the dynamics of the carrier distribution function $f_{n\vk}$. A discussion of these additional equations follows in the next section. 

\section{DEMAGNETIZATION DUE TO CARRIERS IN A  SINGLE SPIN-SPLIT BAND} \label{sec:1band}
Let us concentrate now on a model of a single spin-split band of s symmetry. 
From it, we deduce the general features of the behavior of the system in a simple way. The treatment of the dynamical polarization of carriers is especially transparent in this case. We make use of the distribution functions $f_{s}$ for spin $s$$=$$\uparrow$,$\downarrow$ characterized by two different chemical potentials $\mu_{s}$ and a common temperature $T_{\text{e}}$ (see Sec. \ref{sec:bath}).

We define reduced transition rates $W_{+-}$ and $W_{-+}$ given by $W_{m,m-1}$ and $W_{m,m+1}$, respectively, with the localized spin matrix elements $S^{\pm}_{m,m\mp 1}$ removed (see Eq.~(\ref{eq:Wmn})).  
In a single band with spin-splitting $\Delta$, the transition rate $W_{+-}$ can be rewritten using the spin-resolved densities of states $D_{s}(E)$:
\begin{eqnarray}
W_{+-}  & = &  \frac{\gamma^{2}}{4} \frac{2 \pi}{\hbar}   \int  dE \,   f_{\uparrow}(E)(1-f_{\downarrow}(E-\delta)) \nonumber \\ 
& & \times \, D_{\uparrow}(E) D_{\downarrow}(E-\delta) \,\, ,
	\label{eq:W1}
\end{eqnarray}
and $W_{-+}$ is obtained by exchanging the spins and changing the sign of $\delta$.
Manipulating the explicit forms of occupation functions we obtain a generalization of the detailed balance condition to the case of different chemical potentials for two spin directions:
\begin{equation}
\frac{W_{-+}}{W_{+-}} = e^{\beta_{\text{e}} ( \delta - \Delta \mu)}  \,\, , \label{eq:balance}
\end{equation}
where $\beta_{\text{e}}$$=$$1/k_{B}T_{\text{e}}$ and $\Delta \mu$$=$$\mu_{\uparrow}-\mu_{\downarrow}$ is the spin splitting of the carriers' chemical potential. For $\mu_{\uparrow}$$=$$\mu_{\downarrow}$ we recover the usual detailed balance condition. 

When $S$$=$$1/2$, Eq.~(\ref{eq:general_dSdt}) can be transformed into the Bloch-like equation for the dynamics of average localized spin $\langle S_{z}(t) \rangle$:
\begin{equation}
\frac{d}{dt} \langle S_{z}(t) \rangle = - \frac{ \langle S_{z}(t) \rangle - S_{0}(t)}{T_{1}(t)} \,\, , \label{eq:Bloch}
\end{equation}
where $S_{0}(t)$ is the {\it instantaneous} equilibrium value of the spin, given by the transitions rates at time $t$:
\begin{equation}
S_{0} = \frac{1}{2} \frac{ W_{+-} - W_{-+}}{ W_{+-} + W_{-+}} = -\frac{1}{2}\tanh (\beta_{\text{e}}(\delta - \Delta \mu) ) \,\, ,
\end{equation}
and the relaxation time is given by
\begin{equation}
T_{1}(t) = ( W_{+-}+W_{-+} )^{-1} \,\, .  \label{eq:T1}
\end{equation}

Note that in applications to (III,Mn)V semiconductors we are going to be interested in the regime of $\beta_{\text{e}}\delta$$\ll$$1$ and in the localized spin $S$$=$$5/2$. 
If the dynamic spin splitting also fulfills $\beta_{\text{e}}\Delta \mu$$\ll$$1$, we can approximate Eq.~(\ref{eq:general_dSdt}) {\it for any magnitude of spin S} by the following expression:
\begin{equation}
\frac{d}{dt} \langle S_{z}(t) \rangle \simeq -2W_{+-}(t) \langle S_{z}(t) \rangle \,\, . \label{eq:dSdt_approx}
\end{equation}
For the conditions considered below, this equation gives a very good description of the initial stage of the localized spin dynamics, in which the carrier temperature is very high.
When the temperature drops so that the above inequalities are violated, one has to solve the full eqs.~(\ref{eq:rate}) and (\ref{eq:general_dSdt}) for $S$$>$$1/2$, and the Bloch equation (\ref{eq:Bloch}) for $S$$=$$1/2$.

We now introduce the phenomenological equations describing the dynamics of the carrier bath. The time dependence of $T_{\text{e}}$ comes from the cooling of carriers by phonon emission, and we model it by a simple decay (see Section \ref{sec:bath})
\begin{equation}
T_{\text{e}}(t) = T_{\text{e}}(0) \, e^{-t/\tau_{\text{E}}} \,\, ,	\label{eq:TC}
\end{equation}
where $\tau_{\text{E}}$ is the energy relaxation time of highly excited carriers. The changes of chemical potentials are governed by a second phenomenological equation for the dynamics of the average carrier spin $\langle s(t) \rangle$:
\begin{equation}
\frac{d}{dt}  \langle s(t) \rangle= -\frac{n_i}{n_c} \frac{d}{dt} \langle S(t) \rangle- \frac{ \langle s(t) \rangle - s_{0}(\Delta,T_{\text{e}}) } {\tau_{\text{sr}}} \,\, . \label{eq:dsdt}
\end{equation}
The first term on the right describes the transfer of angular momentum by spin-flips (with $n_{c}/n_{i}$ being the ratio of the carrier density to the localized spin density), and the second term describes the relaxation (on time scale $\tau_{\text{sr}}$) of the average carrier spin towards the instantaneous equilibrium value $s_{0}$ determined by the  spin splitting $\Delta$ and the carrier temperature $T_{\text{e}}$. 
The above set of single-band equations was used in Refs. \onlinecite{Wang_PRL05} and \onlinecite{Wang_JPC06}  to qualitatively model the demagnetization in (III,Mn)V semiconductors.\cite{Wang_PRL05,Wang_JPC06}

\subsection{The inverse Overhauser effect}
The demagnetization process described by the above equations occurs in the following way. We model the absorption of a $\sim$$100$ fs light pulse as an instantaneous increase of concentration and temperature of carriers.
The heating of the carriers by a pulse of light modifies the spin-flip transition rates $W_{+-}$ and $W_{-+}$. The broader the carrier distributions are (the higher the $T_{\text{e}}$ is), the larger these rates are. When the dynamic spin splitting $\Delta \mu$ is zero, the detailed balance of transition rates tells us that the localized spin is going to evolve towards a new value corresponding to a high temperature $T_{\text{e}}$. This final value of $\langle S_{z} \rangle$, and the rate at which it is approached, change with the decrease of the carrier temperature. In addition, the possible build-up of polarization of the carriers changes the spin-flip transition rates in such a way that  the spin transfer is blocked, and without spin relaxation in the carrier system a ``polarization bottleneck'' occurs. This is analogous to the  ``magnetic resonance bottleneck''  known in electron spin resonance of localized moments in metals.\cite{Orbach_JMMM80} In the latter, the resonance spectrum of the localized spins is changed when the spin relaxation of carriers is long enough for the two spin systems to become ``locked'' together in precession.
Here, in the extreme case of very slow spin relaxation, the initial demagnetization can result in a flip of all the carriers of one spin direction, leaving the carrier system in a ``dynamic half-metallic state''. After such a saturation of demagnetization process, the rate of spin flip is determined by the (low) rate of the carrier spin relaxation.

The basic principle of demagnetization is analogous to the well-known Overhauser effect,\cite{Abragam} in which the itinerant spins are optically pumped, and this injected polarization is transferred by s-d type interaction to the localized spins. Although the original Overhauser effect involves pumping of angular momentum into one of the spin populations, the essence of the effect is taking one spin population out of equilibrium with another and thus inducing the transfer of angular momentum between them. One generalization of the Overhauser effect was proposed and realized experimentally by Feher. \cite{Feher_PRL59,Clark_PRL63} The idea was to heat up the electrons by passing a current through the sample in order to induce spin-flips between the electrons spins and nuclear spins. Depending on the parameters of both spin systems, increase of either nuclear spin polarization \cite{Feher_PRL59} or electron polarization \cite{Suhl_arxiv02} is predicted. The ``inverse Overhauser effect'' presented here is related to the latter: by heating up the carriers, we induce the transfer of angular momentum from localized spins to electron (hole) spins.

\begin{figure}
\includegraphics[width=8cm]{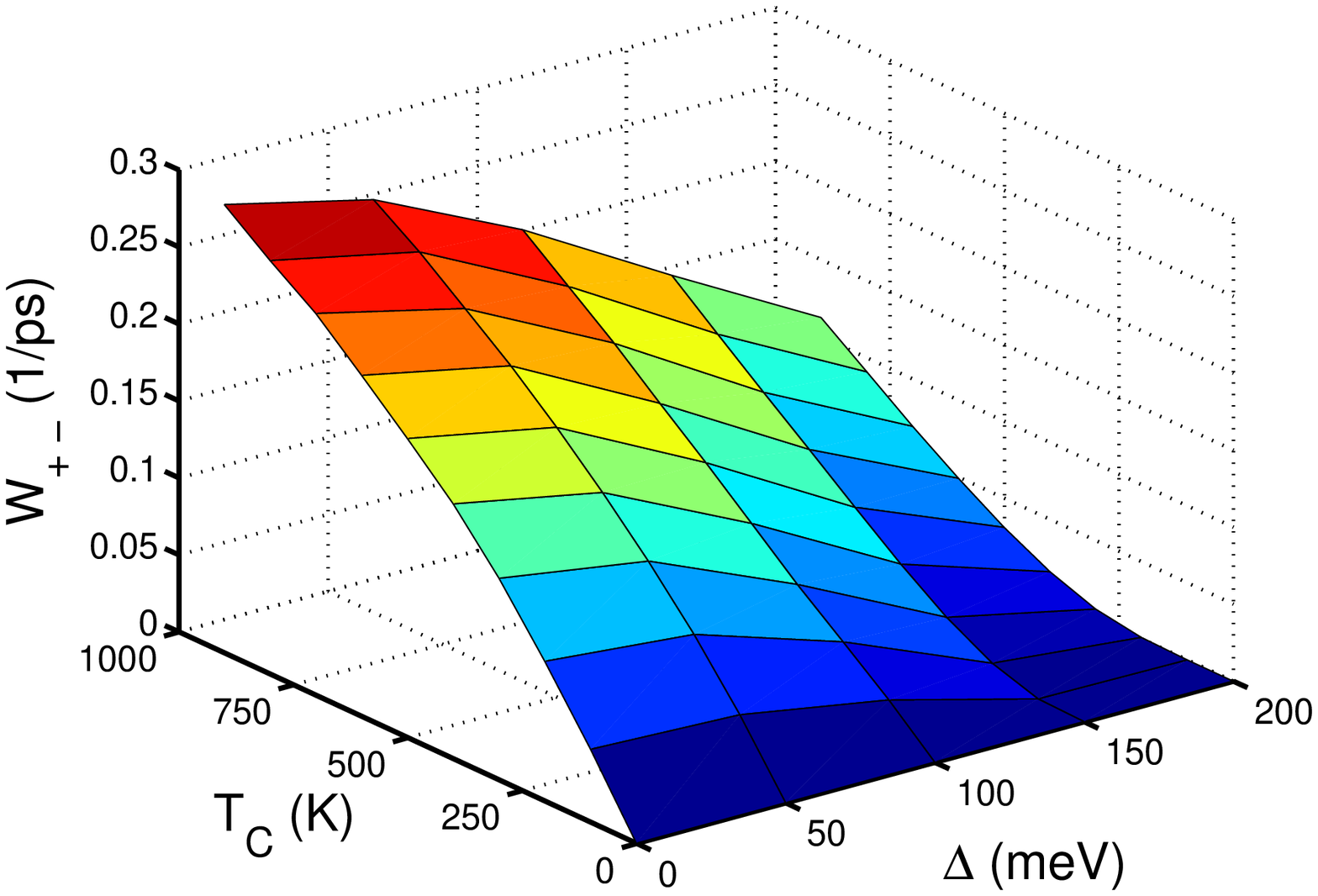}
\caption{(Color online) Transition rate $W_{+-}$ in a single parabolic band with effective mass $m_{\text{eff}}$$=$$1$ and $\gamma$$=$$50$ meV$\cdot$nm$^{3}$ as a function of carrier temperature and band splitting $\Delta$. 
The concentration of carriers is $10^{20}$ cm$^{-3}$.} \label{fig:W1_TD}
\includegraphics[width=8cm]{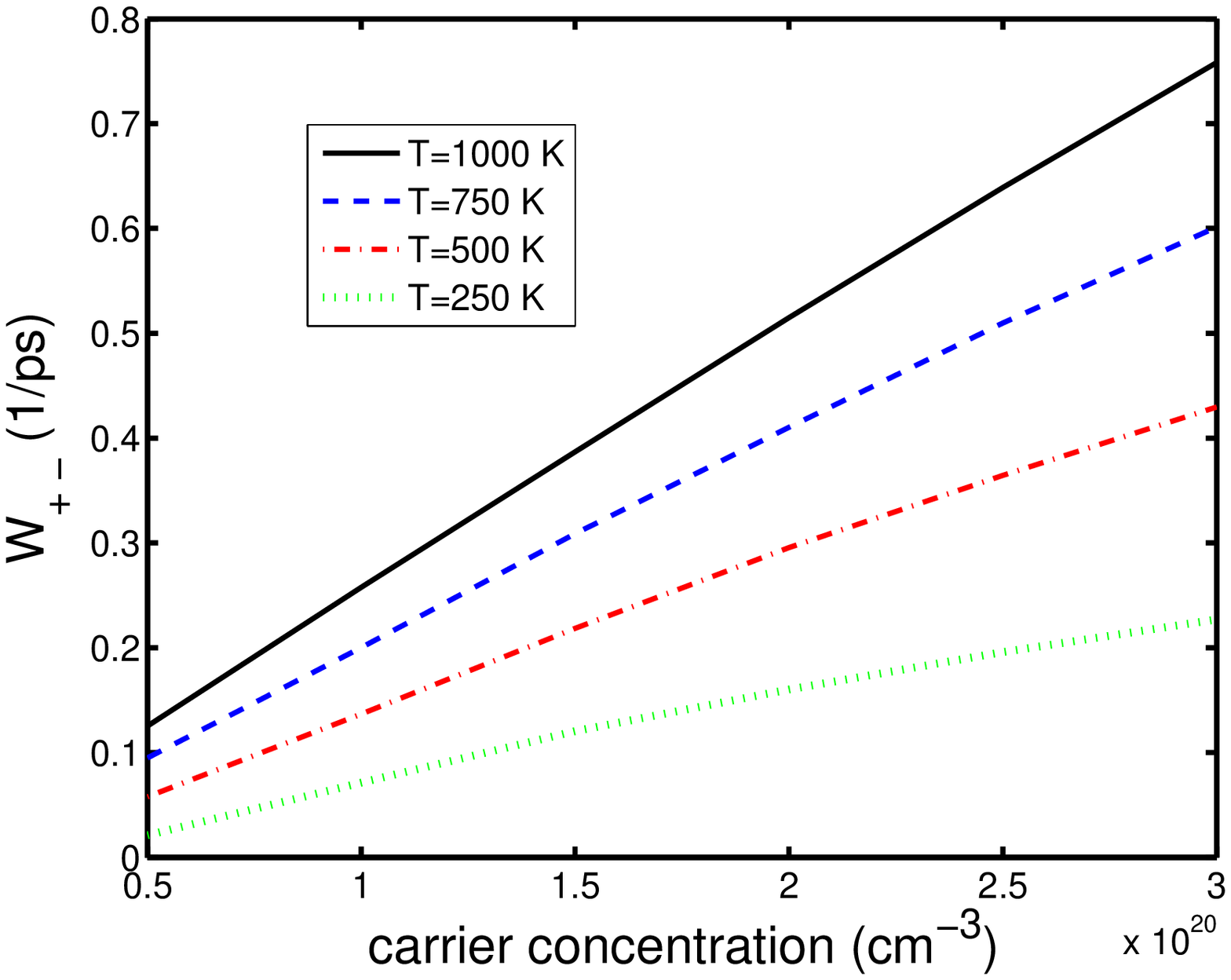}
\caption{(Color online) Transition rate $W_{+-}$ for $\Delta$$=$$0.15$ eV and different carrier temperatures, as a function of carrier density. Other parameters are the same as in Figure \ref{fig:W1_TD}} \label{fig:W1_Tp}
\end{figure}

\begin{figure}
\includegraphics[width=8cm]{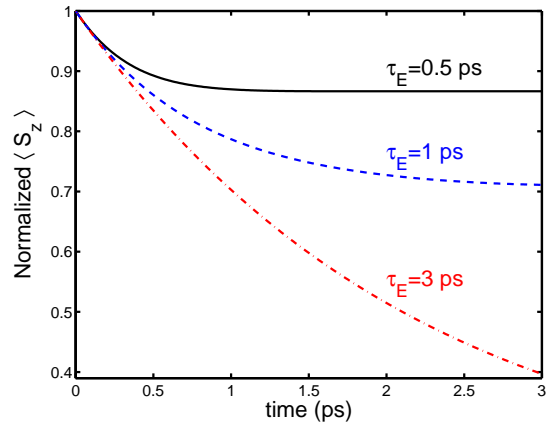}
\caption{(Color online) Demagnetization of the localized spins S$=$$5/2$ (normalized to 1) in a single band model with $m_{\text{eff}}$$=$$1$ and $\gamma$$=$$50$ meV$\cdot$nm$^{3}$ 
for different energy relaxation times of carriers. The concentration of carriers $n_{c}$$=$$10^{20}$ cm$^{-3}$, and the density of localized spins is such that the band splitting  $\Delta$$=$$150$ meV (corresponding to $x$$\approx$$0.055$ in ferromagnetic semiconductors).
} \label{fig:demag1_dmu0}
\end{figure}

\subsection{Carriers as a reservoir of angular momentum: $\Delta \mu$$=$$0$}  \label{sec:dmu0}
We first analyze the case in which the carrier spin relaxation is so fast that the carriers are a good sink of polarization, so that $\Delta \mu$$=$$0$ at each moment of time. The occupation factors $f_{s}$ are the same for both spins, and are characterized by the time varying temperature $T_{\text{e}}$ and chemical potential $\mu$.

When the temperature is not too high, i.e., $k_{B}T_{\text{e}}$ is smaller than the energy scale on which the densities of states change appreciably (but still much larger than localized spin splitting $\delta$), the transition rate (\ref{eq:W1}) can be approximated by
\begin{equation}
W_{+-}(t) \approx \frac{\gamma^{2}}{4} \frac{2 \pi}{\hbar} k_{B}T_{\text{e}}(t) D_{+}(\mu)D_{-}(\mu) \,\, . \label{eq:Wapp}
\end{equation}
It shows that the rate of demagnetization scales with $T_{\text{e}}$ and that large densities of states for both spins around the Fermi level are needed. Together with the $\gamma^{2}$ scaling, this shows that in (III,Mn)V ferromagnetic semiconductors the  holes (having larger mass and exchange constant) are much more effective in the demagnetization process than the electrons. 

Fig.~\ref{fig:W1_TD} and \ref{fig:W1_Tp} illustrate the dependence of $W_{+-}$ from Eq.~(\ref{eq:W1}) on temperature $T_{\text{e}}$, spin splitting $\Delta$, and carrier concentration. We use the density of states of a parabolic band, with effective mass $m_{\text{eff}}$$=$$1$, which roughly corresponds to the density-of-states mass in the valence band of GaMnAs. $W_{+-}$ goes to zero for low carrier temperature, when phase-space blocking limits the number of states which can scatter. It also decreases for increasing spin-splitting $\Delta$, when the number of minority spins available for spin-flip goes down. This effect is stronger for smaller carrier concentrations and lower temperatures.
For $T_{\text{e}}$ close to 1000 K, the corresponding  characteristic time $T_{1}$ is of the order of a picosecond.
Whether a full demagnetization occurs depends on the rate of the energy relaxation of carriers. If the carrier temperature does not drop significantly within the time  $T_{1}(0)$ (calculated at the initial carrier temperature $T_{\text{e}}$), then a significant demagnetization occurs on this time-scale. On the other hand, if the $T_{\text{e}}$ changes strongly on the scale of $T_{1}(0)$, then we have to solve our equations with $W_{+-}(t)$ updated according to carrier temperature changes from Eq.~(\ref{eq:TC}). We are interested in the dynamics occurring during the first picosecond. When  $T_{1}(t)$ becomes much larger than 1 ps, then for our purposes the demagnetization process is effectively stopped. Such an  effect of cooling of carriers on demagnetization is illustrated in Figure \ref{fig:demag1_dmu0}. For all energy relaxation times $\tau_E$ the initial slope of $\langle S_{z}(t) \rangle$ is the same, given approximately by Eq.~(\ref{eq:Wapp}) evaluated for $T_{\text{e}}(0)$, but the time at which the demagnetization ceases and the saturation value depend on  $\tau_E$. This shows that the time-scale on which the ultrafast demagnetization occurs can be given by $\tau_{\text{E}}$, which is not related to magnetic properties of the material. From this point of view, it is not the fact that the magnetization drop occurs in less than a picosecond which is interesting. Instead it is the {\it magnitude} of the demagnetization which demands explanation. 

\subsection{The effects of the dynamic spin polarization of carriers: $\Delta \mu$$\neq$$0$}  \label{sec:bottleneck}

\begin{figure}
\includegraphics[width=8cm]{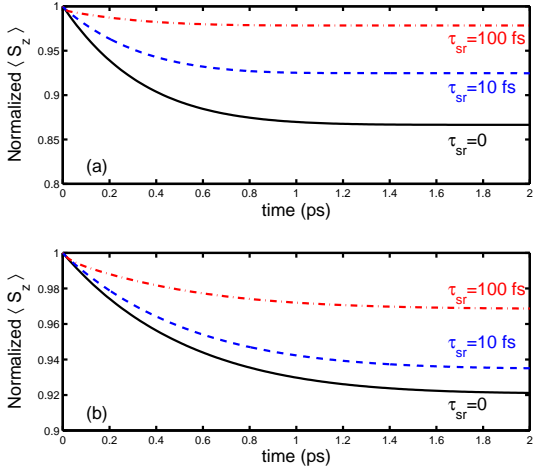}
\caption{(Color online) Demagnetization of the localized spin S for different carrier spin relaxation times, with energy relaxation time $\tau_{\text{E}}$$=$$0.5$ ps. The effective mass of a single spin-split band is a) $m_{\text{eff}}$$=$$1$ and b) $m_{\text{eff}}$$=$$0.5$. All the other parameters are same as in Fig.~\ref{fig:demag1_dmu0}. The total drop in magnetization becomes smaller for longer carrier spin relaxation times.}
\label{fig:bottleneck}
\end{figure} 

The efficiency of the ``inverse Overhauser effect'' is limited by the finite spin relaxation time of the carriers.
Each spin flip which leads to the demagnetization, if not followed by carrier spin relaxations, diminishes the phase space available for next transition of this kind. 
The result is the decrease of the net number of spin-flips during the time $\tau_{\text{E}}$, which translates into smaller total demagnetization. 

Let us concentrate on the case in which $W_{+-}$ is the transition which leads to the demagnetization of the localized spins. The corresponding electron spin flip is from spin $\uparrow$ to $\downarrow$, so that the demagnetization of S spins leads to $\Delta \mu$$<$$0$. In the limit of negligible $\delta$ and $\Delta \mu$ smaller than $k_{B}T_{\text{e}}$, we can approximate Eq.~(\ref{eq:Wapp}) by
\begin{eqnarray}
W_{+-}(t) & \approx & \frac{\gamma^{2}}{4} \frac{2 \pi}{\hbar} k_{B}T_{\text{e}}(t) D_{+}(\mu_{\downarrow})D_{-}(\mu_{\downarrow}) \nonumber \\
& & \times  \Big( 1 +\frac{\Delta \mu}{2k_{B}T_{\text{e}}} + ... \Big) \,\, .	\label{eq:Wapp_dmu}
\end{eqnarray}
From this we see that $W_{+-}$ decreases in the presence of dynamic spin polarization. When $|\Delta \mu|$ becomes comparable to $k_{B}T_{\text{e}}$, $W_{+-}$ goes to zero. 

A rough estimate of what $\tau_{\text{sr}}$ is short enough to be considered instantaneous can be given assuming a constant density of states $D$ for both spins. The localized spin splitting is then $|\delta|$$=$$|\gamma\Delta|D/2$, and the second term in Eq.~(\ref{eq:dsdt}), corresponding to spin relaxation can be written as $D\Delta \mu / 2n_{c}\tau_{\text{sr}}$. We want it to dominate over the first term (spin transfer from localized S) for $\Delta \mu$ small enough, that the transition rate $W_{+-}$ is still unaffected by such dynamic spin splitting, i.e.~for $\Delta \mu$$\ll$$k_{B}T_{\text{e}}$. The resulting inequality is
\begin{equation}
\frac{D k_{B} T_{\text{e}}}{4 S n_{i}} > W_{+-}(\Delta \mu \approx 0, T_{\text{e}}) \tau_{\text{sr}} \,\, ,
\end{equation}
which means that the ratio of the density of carriers available for the spin flip to the localized spin density is larger than the product of spin-flip rate and carrier spin relaxation time. Using Eq.~(\ref{eq:Wapp}) we can transform this inequality into a very simple, but physically less intuitive form:
\begin{equation}
\tau_{\text{sr}} < \frac{\hbar}{4\pi \delta} \,\, ,  \label{eq:bottleneck_inequality}
\end{equation}
In (III,Mn)V semiconductors for typical value of localized spin splitting $\delta$ is of the order of 1 meV, the spin relaxation time has then to be smaller than 100 fs for spin bottleneck to become unimportant.

The ``spin bottleneck'' effect is illustrated in Figure \ref{fig:bottleneck}, where calculated $\langle S_{z}(t) \rangle$ are plotted for different values of carrier spin relaxation time $\tau_{\text{sr}}$. The bottleneck effect is stronger in Figure \ref{fig:bottleneck}a, where the effective mass $m_{\text{eff}}$$=$$1$ and carrier concentration $n_{c}$$=$$10^{20}$ cm$^{-3}$. In Figure \ref{fig:bottleneck}b, where $m_{\text{eff}}$$=$$0.5$, the difference between results for $\tau_{\text{sr}}$$=$$10$ fs and $\tau_{\text{sr}}$$=$$0$ is smaller. When the effective mass $m_{\text{eff}}$ is smaller, while keeping $n_{c}$ and $\Delta$ the same, the average carrier spin is decreased, as the Fermi energy becomes larger compared with $\Delta$. Then $\delta$ is smaller, making the inequality (\ref{eq:bottleneck_inequality}) easier to fulfill.

\section{DEMAGNETIZATION DUE TO HOLES IN THE VALENCE BAND OF (III,Mn)V SEMICONDUCTOR} \label{sec:valence}
For the case of (III,Mn)V semiconductors the carriers relevant for ultrafast demagnetization are the holes. If they reside in an impurity band, the single band theory described above could be applied, but currently there is no simple quantitative model for this case. Below we perform calculations for the case of holes residing inside the valence band of a bulk semiconductor.  We use an ``effective Hamiltonian'' model \cite{Dietl_PRB01,Jungwirth_RMP06} in which a mean-field p-d term from Eq.~(\ref{eq:Cmf}) is added to the 6 band Luttinger Hamiltonian. In this way we can analyze quantitatively the influence of the strong spin-orbit interaction on the spin-flip transition rate.

\begin{figure}
\includegraphics[width=8cm]{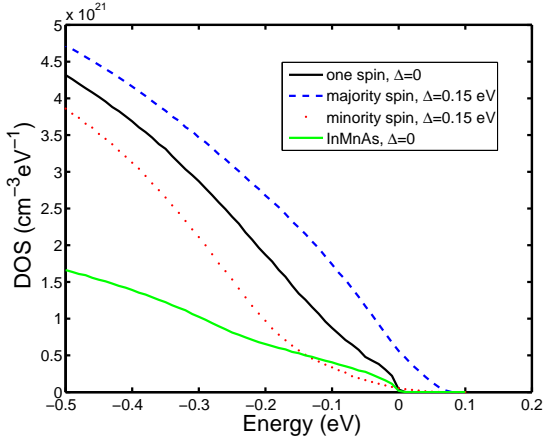}
\caption{(Color online) Valence band spin-resolved densities of states in GaMnAs. The solid lines are for single spin with $\Delta$$=$$0$ in GaMnAs and InMnAs. Dashed and dotted  lines are majority and minority spin for  splitting $\Delta$$=$$0.15$ eV in GaMnAs. }
\label{fig:DOS_dms}
\end{figure}

A typical plot of energy dispersions in (III,Mn)V calculated by $\mathbf{k}\cdot\mathbf{p}$ method with exchange splitting $\Delta$$=$$0.15$ eV is shown in Figure \ref{fig:GaMnAs_bandstructure}. 
The corresponding spin-resolved densities of states are shown in Figure  \ref{fig:DOS_dms}. 
In order to calculate the $W_{+-}$ transition rate we cannot use Eq.~(\ref{eq:W1}), which requires only the densities of states. 
In the p symmetry band the matrix elements of carrier ladder operators $\hat{s}^{\pm}$ are non-trivial due to spin-orbit coupling, and cannot be neglected. The full Eq.~(\ref{eq:Wmn}) has to be used. 

\begin{figure}
\includegraphics[width=8cm]{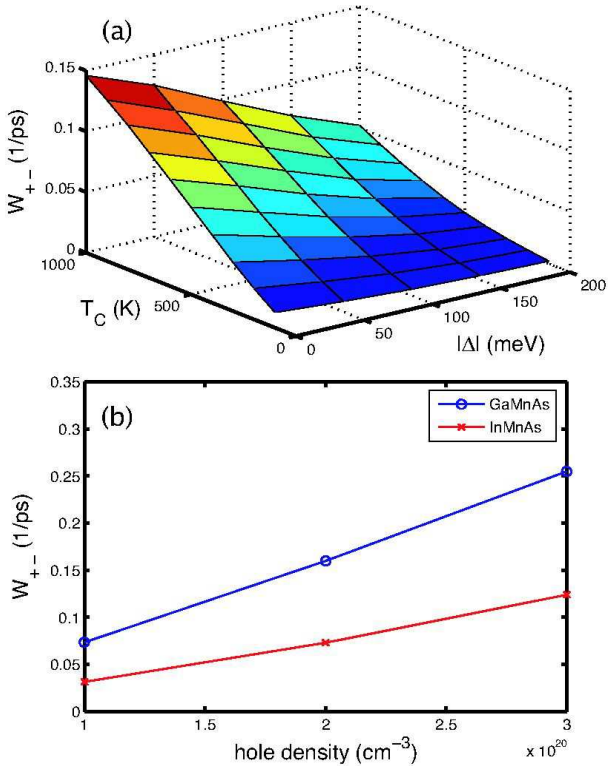}
\caption{(Color online) (a) Reduced transition rate $W_{+-}$ in GaMnAs for hole density $p$$=$$10^{20}$ and no dynamic polarization of holes ($\Delta\mu$$=$$0$). Analogous plots for other values of $p$ and for InMnAs look qualitatively similar, only the overall magnitude of $W_{+-}$ changes. (b) Hole density dependence of $W_{+-}$ at $T_{\text{e}}$$=$$1000$ K and spin splitting $\Delta$$=$$-0.15$ eV (the lines are guides for the eye).}
\label{fig:W6x6}
\end{figure}

Strong spin-orbit interaction suppresses the spin-flip transition rate, compared to the value predicted by Eq.~(\ref{eq:W1}). If we disregard for a moment the orbital parts of the $|n\vk\rangle$ eigenstates, the squared matrix element $|\langle n'\vk'|\hat{s}^{-}|n\vk \rangle |^{2}$ is maximal when the spin part of $|n\vk\rangle$ state is $| \! \uparrow\rangle$, and the spin part of  $|n'\vk'\rangle$ is $| \! \downarrow\rangle$. If both states contained equal mixtures of spin up and down, the squared matrix element would be $1/4$. When the orbital parts of the states are taken into account, the spin-orbit interaction diminishes the matrix elements between the states with different spin character, because it aligns orthogonal orbital wave functions (of different orbital angular momentum $l$) with opposite spins $s$. 
In the limit of infinite spin-orbit interaction, a spin-flip between pure spin-up and spin-down states is impossible, because the orbital parts of $|n\vk\rangle$ states are exactly orthogonal. This limit is realized in the case of the 4 band Luttinger Hamiltonian for the holes. We have evaluated analytically the spin-flip transition rate in a spherical $4\times 4$ Luttinger model at zero spin splitting, and obtained that the result of the exact Eq.~(\ref{eq:Wmn}) is smaller by a factor of $5/18$ than the value obtained from densities of states in Eq.~(\ref{eq:W1}). This sets a limit on how much the spin-orbit interaction can suppress the transition rates for small $\Delta$ in a full $6$ band model.

For the actual calculation of the time-dependence of magnetization, we have evaluated the transition rates from  Eq.~(\ref{eq:Wmn}) using the band-structure obtained from the $6$$\times$$6$ Luttinger Hamiltonian. 
In all the following calculations, we have put $\delta$$=$$0$, which is justified by the smallness of $\delta$ in ferromagnetic semiconductors and the fact that we are primarily interested in the regime of high $T_{\text{e}}$, where $\delta$ can be completely disregarded.
In Figure \ref{fig:W6x6} we plot $W_{+-}$ for GaMnAs and InMnAs for various carrier temperatures, exchange splittings and hole densities. 
The transition rates are larger for GaMnAs, which can be traced mostly to larger density of states in the valence band compared to InMnAs. 

The presence of strong spin-orbit interaction makes including the dynamic spin polarization of carriers much harder than in a single-band case. 
Due to a large spin-orbit splitting $\Delta_{\text{SO}}$$\simeq$$0.3-0.4$ eV the $|n\vk\rangle$ states have mixed spin character even in presence of exchange splitting $\Delta$$\approx$$0.1$ eV, and a simple introduction of different distribution functions $f_{s}$ for two spin directions is not possible. Performing an ensemble Monte Carlo calculation of dynamics of strongly excited holes\cite{Bailey_PRB90} with spin degree of freedom taken into account  is a vast undertaking beyond the purpose of this investigation. We resort to an approximate method of including the effects of non-zero  $\Delta \mu$. 

\begin{figure}
\includegraphics[width=7cm]{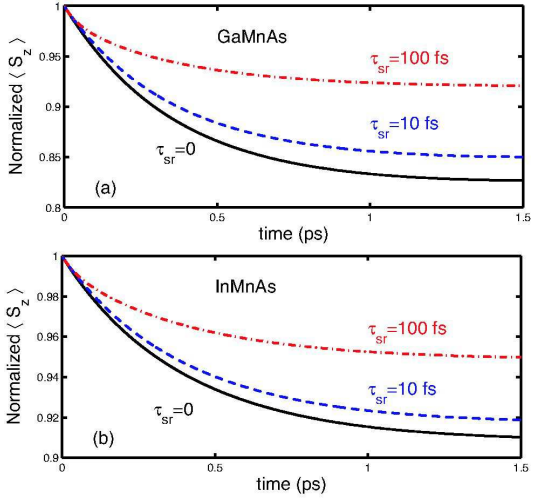}
\caption{(Color online) Demagnetization of the Mn spin in (a) GaMnAs and (b) InMnAs. The solid lines are calculated using the exact transition rates at $\Delta \mu$$=$$0$ (dynamical spin polarization of carriers is neglected). The dashed lines come from the approximate calculation including the finite hole spin relaxation time $\tau_{\text{sr}}$ (see text for details). The initial temperature of holes if $T_{\text{e}}$$=$$1000$, the hole density is $p$$=$$3\times 10^{20}$ cm$^{-3}$, the exchange integral $\beta$$=$$-50$ meV$\cdot$nm$^{3}$, and the molar fraction of Mn is $x$$=$$0.054$ (initial spin splitting $\Delta$$=$$-0.15$ eV). The energy relaxation time $\tau_{\text{E}}$ is $0.5$ ps.}
\label{fig:demag6x6}
\end{figure}

Above we have calculated the exact transition rates for $\Delta \mu$$=$$0$. We have also calculated the corresponding $W_{+-}$ assuming constant matrix elements and only taking spin-resolved densities of states calculated from Luttinger Hamiltonian (Eq.~(\ref{eq:W1})). The inclusion of matrix elements leads to decrease of $W_{+-}$ by a factor of $0.25$-$0.3$ in the most relevant range of high $T_{\text{e}}$ and $\Delta$$\simeq$$0.1$ eV (not shown). We assume that this ratio of exact and DOS-derived $W_{+-}$  also holds in the case of a finite $\Delta \mu$ (but small compared to $E_{F}$ and $k_{B}T_{\text{e}}$). We have calculated the demagnetization with finite spin relaxation time of the holes using the properly down-scaled transition rates from Eq.~(\ref{eq:W1}) with finite $\Delta \mu$. The results are shown in Figure \ref{fig:demag6x6}. The effect of spin bottleneck on demagnetization is clearly very weak for $\tau_{\text{sr}}$$\leq$$10$ fs. For $\tau_{\text{sr}}$$=$$100$ fs the total demagnetization is diminished by $\sim$$50$\%. These results agree with estimates from the inequality (\ref{eq:bottleneck_inequality}), which is fulfilled for $\delta$$\sim$$1$ meV and $\tau_{\text{sr}}$$=$$10$ fs. 

In the presence of the  strong spin-orbit interaction, the hole spin relaxation is expected to be very fast, occurring on the momentum scattering time-scale.\cite{Optical_Orientation} For example, in pure GaAs the spin relaxation of holes was measured to be about 100 fs.\cite{Hilton_PRL02}  The calculated \cite{Jungwirth_APL02} momentum scattering of holes in GaMnAs at low temperature is about 10 fs, and it is expected to be shorter when the holes are highly excited. There is one caveat in the case of exchange spin-split bands in (III,Mn)V. The lifting of the heavy-light hole degeneracy by confinement in quantum wells makes the spin relaxation time longer,\cite{Uenoyama_PRL90,Ferreira_PRB91} with measured values of  4 ps for holes close to $\Gamma$ point in modulation-doped GaAs quantum wells.\cite{Damen_PRL91}
An analogous effect is expected in exchange-split bands for holes with small wave-vectors.
However, for the carrier densities and temperatures under consideration we are mostly interested in spin relaxation of holes with quite large wavevectors ($|\mathbf{k}|$$\approx$$1$$-$$3$ nm$^{-1}$). 
For such large $|\mathbf{k}|$ the effect of lifted degeneracy on  spin relaxation becomes much weaker.\cite{Ferreira_PRB91} For example, at a wavevector of this magnitude in the direction perpendicular to the magnetization the band-mixing terms in the Luttinger Hamiltonian overcome the p-d interaction, and the spin splitting becomes negligible. 
Taking all this into account it is reasonable assume that in an excited and disordered sample the hole spin relaxation time should be shorter than in pure GaAs. For  $\tau_{\text{sr}}$$<$$100$ fs we can see in  Fig.~\ref{fig:demag6x6} that the spin bottleneck effect is qualitatively unimportant, and it becomes negligible for $\tau_{sr}$$=$$10$ fs. 

In the experiments on (III,Mn)V spin lifetimes of electrons between $~$1 ps in highly excited InMnAs\cite{Wang_PE} and 30 ps in much more weakly excited GaMnAs\cite{Kimel_PRL04} were seen. To the best of our knowledge, no signal attributable to hole spin relaxation has ever been seen on a time-scale of the temporal resolution of the experiments. If $\tau_{\text{sr}}$$\leq$$10$ fs, the measurements would be particularly challenging due to time-resolution constraints on the laser pulse time. Also, at such time-scale the carrier dynamics is in the non-Markovian regime,\cite{Wegener_PRL98} and $\tau_{\text{sr}}$ is a typical time-scale of the spin-dependent correlated relaxation dynamics.

In Fig.~\ref{fig:demag6x6} we can see that for $\tau_{\text{sr}}$$<$$100$  the spin bottleneck effect is qualitatively unimportant, and it becomes negligible for $\tau_{\text{sr}}$$=$$10$ fs. Without the bottleneck, we obtain a typical magnitude of demagnetization in (III,Mn)V within a picosecond to be of the order of 10\%. For the same carrier density, initial $T_{\text{e}}$$=$$1000$ K and $\tau_{\text{E}}$$=$$0.5$ ps, the drop in magnetization is more pronounced in GaMnAs ($20$\% demagnetization)  than in InMnAs ($10$\%). A discussion of the connection between these calculations and experiments follows below. 

\section{Comparison with experimental results} \label{sec:comparison}
We can estimate the initial temperature of carriers in experiments on InMnAs from Ref.~\onlinecite{Wang_PRL05}  in the following way. There are two time-scales in the demagnetization: an ultrafast one ($<$$1$ ps) and a subsequent much slower ($\sim$$100$ ps) demagnetization, due to spin-lattice relaxation. For example, for pump fluence of $3$ mJ$/$cm$^{2}$, an ultrafast quenching of $\sim 50$\% of magnetization was followed by a complete demagnetization 100 ps later. On the latter time-scale we can safely use a thermal description of the spin system. The full demagnetization means that the localized spin temperature $T_{\text{s}}$ had risen above the Curie temperature, which was 50 K in the sample used. This tells us that 100 ps after the excitation the lattice temperature $T_{\textit{l}}$ was at least 50 K.  The energy (per atom), which had to be transferred into the lattice to heat it up from initial temperature $T_{i}$ (before the excitation) up to the final $T_{\textit{l}}$ is given by
\begin{equation}
\Delta E_{\textit{l}} = \int_{T_{i}}^{T_{\textit{l}}} c_{L}(T) dT \,\, , \label{eq:dEL}
\end{equation}
where $c_{\textit{l}}$ is the specific heat of the lattice (per atom), which in the temperature range of interest is given by the Debye formula:
\begin{equation}
c_{\textit{l}}(T) = \frac{12 \pi^{4}}{5} k_{B} \Big( \frac{T}{\Theta_{D}} \Big) ^{3} \,\, , 
\label{eq:cL}
\end{equation}
where $\Theta_{D}$$=$$280$ K is the Debye temperature of InAs. 

The energy transferred into the lattice needs now to be related to the energy deposited initially by the light pulse into the valence band. In the process of absorption, the energy of the pump photon ($\hbar \omega$$=$$0.6$ eV) is split into a kinetic energy of a created hole and an energy of an excited state within a gap (we disregard the photoelectrons because the low density of states in the conduction band makes this excitation channel insufficient to explain the experiments). The Fermi level before the excitation is situated between $0.1$ and $0.3$ eV below the top of the valence band for hole densities of $10^{20}$ and $3\times 10^{20}$ cm$^{-3}$, respectively. Thus, the energies of the newly created holes are a sizable fraction of the photon energy. As a rather safe estimate we will take the fraction of the pump pulse energy imparted to the holes as $R$$=$$1/4$. 

Now we can derive an estimate of the carrier temperature  $T_{\text{e}}(0)$ just after the absorption of the pump pulse. We assume that after 100 ps all the absorbed energy has been transferred to the lattice.
The observed spin temperature $T_{\text{s}}$ is larger than $50$ K and gives a lower bound for the final lattice temperature $T_{\textit{l}}$. We calculate the energy given to the lattice using Eqs.~(\ref{eq:dEL}) and (\ref{eq:cL}), with initial $T_{i}$$=$$10$ K and final $T_{\textit{l}}$$=$$50$ K.
A fraction $R$ of this energy was an excess kinetic energy of holes in the valence band after the excitation. Using the density of states of InMnAs with $\Delta$$=$$0.15$ eV calculated before, we obtain the temperature of carriers which gives such excess energy. In this way we obtain an estimate of the initial $T_{\text{e}}$, which we find to be between 1500 and 1000 K for hole densities changing between $10^{20}$ and $3\times 10^{20}$ cm$^{3}$. The analogous calculation in GaMnAs gives $T_{\text{e}}$ smaller by a factor of 2 (due to the larger density of states).
These results justify our use of typical initial $T_{\text{e}}$$=$$1000$~K throughout this article.  

Let us now address the corresponding demagnetization in InMnAs for pump fluence of 3 mJ/cm$^{2}$. 
The hole concentration before the excitation is about $3 \times 10^{20}$ cm$^{-3}$, estimated from critical temperature of $50$ K and magnetization measurements.\cite{Wang_PRL05} After the excitation the total number of holes is larger but, because the initial $p$ is not certain, we will simply use a value of total $p$$=$$3 \times 10^{20}$ cm$^{-3}$. 
The measured sample was a 25 nm thick layer of InMnAs grown on GaSb. For such thickness, the confinement leads to formation of hole subbands, with typical energy spacing between them of the order of 10 meV. This confinement energy is much less than the disorder broadening and the thermal spread of photoexcited carriers (both $\sim$$0.1$ eV), so that our calculation of spin-flip transition rates using the bulk band-structure should be a good approximation.
In the previous section, we have seen that for such $p$$=$$3 \times 10^{20}$ cm$^{-3}$ the sub-picosecond drop in magnetization is about $10$\%. Experimentally a $50$\% drop in Kerr signal was observed within $\sim$$200$ fs after the pump pulse. The step-like character of the magnetization drop is not reproduced by our theory, which predicts a smoother demagnetization. However, the carrier-induced artifacts (termed ``dichroic bleaching'' in Ref.~\onlinecite{Koopmans_PRL00}) obscuring the magnetization dependence of the magnetooptical signal are possible at very short time-scale. The magnitude of the total demagnetization is in qualitative agreement with our calculations.

For pump fluence of 10 mJ/cm$^{2}$, for which a complete quenching of ferromagnetic order is observed, the role of multi-photon absorption processes becomes pronounced, and photocarriers are created in a very large region of the Brillouin Zone (as in experiments from Ref.~\onlinecite{Kojima_PRB03}, where a pump with $\hbar\omega$$=$$3.1$ eV was used in GaMnAs). Our $\mathbf{k}\cdot\mathbf{p}$ model with ``rigid'' spin splitting (i.e.~$\mathbf{k}$-independent exchange constant) is not applicable far away from the $\Gamma$ point. Also, for very high excitation the assumption of quasi-thermal distribution of holes can fail. 

In GaMnAs the experiments show a similar behavior \cite{Wang_JPC06} after 1 picosecond, with  $30$\% magnetization drop for fluence of $\sim$$8$ mJ/cm$^{2}$. This value is very close to the predictions of our theory for $p$$=$$3\times 10^{20}$ cm$^{-3}$.
For the same fluence InMnAs was already nearly completely demagnetized.
Although our theory predicts larger demagnetization in GaMnAs than in InMnAs, in order to make meaningful comparisons one has to achieve comparable excitation parameters (total $p$ and $T_{\text{e}}$ after the pulse). Also, the efficiency of excitation of holes can be different in the two materials.

\section{SUMMARY}
We have presented an investigation of the ultrafast demagnetization induced by an incoherent light excitation in a system described by the sp-d model.
The physical picture of demagnetization is very transparent, with sp-d interaction providing a mechanism of spin transfer from the localized spins into the system of excited carriers. This process is closely related to effects known from systems of electronic and nuclear spins coupled by hyperfine contact interaction, and thus we have termed it the inverse Overhauser effect. The demagnetization basically comes from very fast $T_{1}$ relaxation of the localized spins due to sp-d interaction with the hot carriers' spins. 
In a simple one-band model the rate of spin flip scattering of itinerant and localized spins is proportional to the square of the sp-d exchange constant, the product of electronic densities of states at Fermi energy for two spin directions, and the temperature of the carriers. Due to the carrier temperature dependence the rapid demagnetization can occur only for a time of the order of energy relaxation time $\tau_{\text{E}}$. Thus, the sub-picosecond time scale of the demagnetization process simply comes from the characteristic time of carrier-phonon interaction. The total magnitude of demagnetization during the $\tau_{\text{E}}$ time depends only on the spin-flip scattering efficiency if the carrier system is a good spin sink, i.e. the carrier spin relaxation time $\tau_{\text{sr}}$ is shorter than the time-scale on which the dynamic spin polarization of carriers builds up. A long $\tau_{\text{sr}}$ leads to spin bottleneck that suppresses the demagnetization. 
This general theory can be used as an admittedly rough model for demagnetization process in transition metals (treated within s-d model), as it captures all the crucial aspects of the problem. \cite{Koopmans_TAP03}  

A large part of the discussion was aimed at the specific cases of (III,Mn)V semiconductors. There, a non-trivial band structure with spin-orbit interaction has to be taken into account. We have performed the calculations of spin-flip transition rate using an effective $\vk$$\cdot$$\mathbf{p}$ Hamiltonian approach. 
An approximate calculation of the effects of dynamical spin polarization of holes has been used to argue that  the holes can be treated as a perfect sink for the angular momentum transferred from the localized spins. 
A qualitative agreement with demagnetization experiments in these materials was obtained: the theory shows that the magnetization can drop by $\sim$$10$\% within the energy relaxation time of the holes. 
More theoretical and experimental work on electronic and optical properties of (III,Mn)Vs will be necessary in order to better reconcile experiments and theory.

\section*{ACKNOWLEDGEMENTS}
This work is supported by NSF DMR-0325599. We thank H.~Suhl, J.~Fern{\'a}ndez-Rossier, S.~Saikin and K.S.~Burch for stimulating discussions and comments. {\L}C would like to thank J.~Kono for helpful discussions and encouragement.

\end{document}